\documentclass[prd,twocolumn,nofootinbib,showpacs,superscriptaddress,preprintnumbers,amsmath,amssymb]{revtex4-1}
\usepackage[colorlinks=true, pdfstartview=FitV, linkcolor=red, citecolor=blue, urlcolor=black, pdftitle={},pdfauthor={Tomoya Hayata, Yoshimasa Hidaka},pdfsubject={}, pdfkeywords={}]{hyperref}
\usepackage{graphicx}
\usepackage{setspace,bm}
\usepackage{mathtools}
\usepackage{latexsym,amssymb,amsmath,mathrsfs}
\usepackage{color}

\makeatletter
\providecommand \@ifxundefined [1]{%
 \@ifx{#1\undefined}
}%
\providecommand \@ifnum [1]{%
 \ifnum #1\expandafter \@firstoftwo
 \else \expandafter \@secondoftwo
 \fi
}%
\providecommand \@ifx [1]{%
 \ifx #1\expandafter \@firstoftwo
 \else \expandafter \@secondoftwo
 \fi
}%
\providecommand \natexlab [1]{#1}%
\providecommand \enquote  [1]{``#1''}%
\providecommand \bibnamefont  [1]{#1}%
\providecommand \bibfnamefont [1]{#1}%
\providecommand \citenamefont [1]{#1}%
\providecommand \href@noop [0]{\@secondoftwo}%
\providecommand \href [0]{\begingroup \@sanitize@url \@href}%
\providecommand \@href[1]{\@@startlink{#1}\@@href}%
\providecommand \@@href[1]{\endgroup#1\@@endlink}%
\providecommand \@sanitize@url [0]{\catcode `\\12\catcode `\$12\catcode
  `\&12\catcode `\#12\catcode `\^12\catcode `\_12\catcode `\%12\relax}%
\providecommand \@@startlink[1]{}%
\providecommand \@@endlink[0]{}%
\providecommand \url  [0]{\begingroup\@sanitize@url \@url }%
\providecommand \@url [1]{\endgroup\@href {#1}{\urlprefix }}%
\providecommand \urlprefix  [0]{URL }%
\providecommand \Eprint [0]{\href }%
\providecommand \doibase [0]{http://dx.doi.org/}%
\providecommand \selectlanguage [0]{\@gobble}%
\providecommand \bibinfo  [0]{\@secondoftwo}%
\providecommand \bibfield  [0]{\@secondoftwo}%
\providecommand \BibitemOpen [0]{}%
\providecommand \EOS [0]{\spacefactor3000\relax}%
\providecommand \BibitemShut  [1]{\csname bibitem#1\endcsname}%
\let\auto@bib@innerbib\@empty

\newcommand{\ri}{\mathrm{i}}

\newcommand{\cH}{{H}}

\newcommand{\up}{\uparrow}
\newcommand{\down}{\downarrow}
\newcommand{\be}{\begin{equation}}
\newcommand{\ee}{\end{equation}}
\newcommand{\bea}{\begin{eqnarray}}
\newcommand{\eea}{\end{eqnarray}}
\newcommand{\tr}{\mathrm{tr}}

\begin{document}
\preprint{KEK-TH-2247, J-PARC-TH-0223}
\title{Thermalization of Yang-Mills theory in a $(3+1)$ dimensional small lattice system}

\author{Tomoya Hayata}
\affiliation{Department of Physics, Keio University, 4-1-1 Hiyoshi, Kanagawa 223-8521, Japan}
\author{Yoshimasa Hidaka}
\affiliation{KEK Theory Center, Tsukuba 305-0801, Japan}
\affiliation{Graduate University for Advanced Studies (Sokendai), Tsukuba 305-0801, Japan}
\affiliation{RIKEN iTHEMS, RIKEN, Wako 351-0198, Japan}

\date{\today}

\begin{abstract}
We study the real-time evolution of SU($2$) Yang-Mills theory in a $(3+1)$ dimensional small lattice system after interaction quench. 
We numerically solve the Schr{\"o}dinger equation with the Kogut-Susskind Hamiltonian in the physical Hilbert space obtained by solving Gauss law constraints.
We observe the thermalization of a Wilson loop to the canonical state; the relaxation time is insensitive to the coupling strength and estimated as $\tau_{\rm  eq}\sim 2\pi/T$ with temperatures $T$ at steady states.
We also compute the vacuum persistence probability (the Loschmidt echo) to understand the relaxation from the dynamics of the wave function.
\end{abstract}

\maketitle

\section{Introduction}
How an isolated quantum system reaches thermal equilibrium is one of the fundamental problems in modern physics. In particular, the thermalization of a nonabelian gauge theory is important for understanding the nature of quark-gluon plasmas observed in relativistic heavy-ion collision experiments (See Refs.~\cite{Busza:2018rrf,Berges:2020fwq} for recent reviews). The analysis of relativistic heavy-ion collision experiments using a hydrodynamic model implies that hydrodynamics can be applied from a very early stage after the collision ($\sim0.5$ fm/c). However, a microscopic calculation based on kinetic theory shows that the time scale of thermalization with a small QCD coupling $\alpha_S$ is of order $\alpha_S^{-13/5}Q_s^{-1}$,
where $Q_s$ is the characteristic momentum scale of gluons inside the colliding nuclei.
This time scale is orders of magnitude larger than the one expected in hydrodynamic models~\cite{Baier:2000sb,Baier:2002bt}. On the other hand, analyses based on the gauge/gravity duality imply that the thermalization time is the order of the inverse of temperature $1/(\pi T)$ in the large colors and  large 't Hooft coupling limit~\cite{Horowitz:1999jd,Chesler:2009cy,Heller:2011ju,Heller:2012km}. This rapid thermalization has been thought of as the universal property of the strongly coupled gauge theories, and the quark-gluon plasmas produced in relativistic heavy-ion collision experiments are thought of as strongly coupled, while the conventional plasmas are weakly coupled.

Hydrodynamic behavior has also been observed in small systems of $pp$ and $pA$ collisions~\cite{Abelev:2014mda,Sirunyan:2017uyl,Aaboud:2017acw}. 
This observation is unexpected because it is usually thought that thermalization does occur due to a large number of degrees of freedom.
There are two possibilities: One is that this is the property of quantum field theories. 
Since the dimension of the  Hilbert space is infinite, thermalization does occur even in small systems.
The other is that hydrodynamics works well before thermalization where the pressure is isotropic, which is called hydrodynamization. This possibility has been intensively studied in recent years (see Refs.~\cite{Berges:2020fwq,Shen:2020mgh} for recent reviews).

In this paper, we discuss the first possibility of quantum thermalization.
For this purpose, we study the thermalization of the SU(2) Yang-Mills theory in a small isolated system.
 We employ the Kogut-Suskind Hamiltonian formulation on a single cubic lattice with open boundary conditions~\cite{Kogut:1974ag} and numerically solve the Schr{\"o}dinger equation. 
To mimic the situation in heavy-ion collision experiments, we employ interaction quench from the strong gauge coupling limit to a weak gauge coupling. 
This enables us to avoid the aforementioned problem of whether thermalization, hydrodynamization, or isotropization. Furthermore, the thermalization mechanism will be purely quantum because there is no kinetic regime on the small lattice.

The advantage of the Hamiltonian formulation is free from the so-called sign problem in Monte Carlo simulations, that is, the difficulty of importance sampling due to the complexity of the path-integral weight~\cite{Alexandru:2020wrj}.
On the other hand, the disadvantage is the exponentially large Hilbert space. However, as shown below, we overcome this difficulty by considering a small system and 
explicitly solving the Gauss law constraints, which numerously reduces the size of physical Hilbert space, and enables us to access the real-time dynamics of the Yang-Mills theory using the standard classical computers.
Our finding has a substantial impact on developing fields of classical or quantum simulations of lattice gauge theories (see Refs.~\cite{Zohar_2015,Banuls:2019bmf} for review).

This paper is organized as follows.
In Sec.~\ref{sec:Hamitonian formalism}, we briefly review the Kogut-Susskind Hamiltonian formulation and show how to construct the physical space and the matrix element of Hamiltonian.
In Sec.~\ref{sec:Real time simulation}, we discuss the real-time dynamics by solving the Schr{\"o}dinger equation.
We show that the results of our numerical simulations exhibit thermalization, and the thermalization time is of order of the Boltzmann time, $\tau_{\rm eq}\sim 2\pi/T$.
Section~\ref{sec:Summary} is devoted to summary and outlook.
In Appendix~\ref{sec:Eigenvalues of the Hamiltonian}, we show the eigenvalues of the Hamiltonian near the ground state.

\begin{figure}[t]
  \centering
   \includegraphics[width=.25\textwidth]{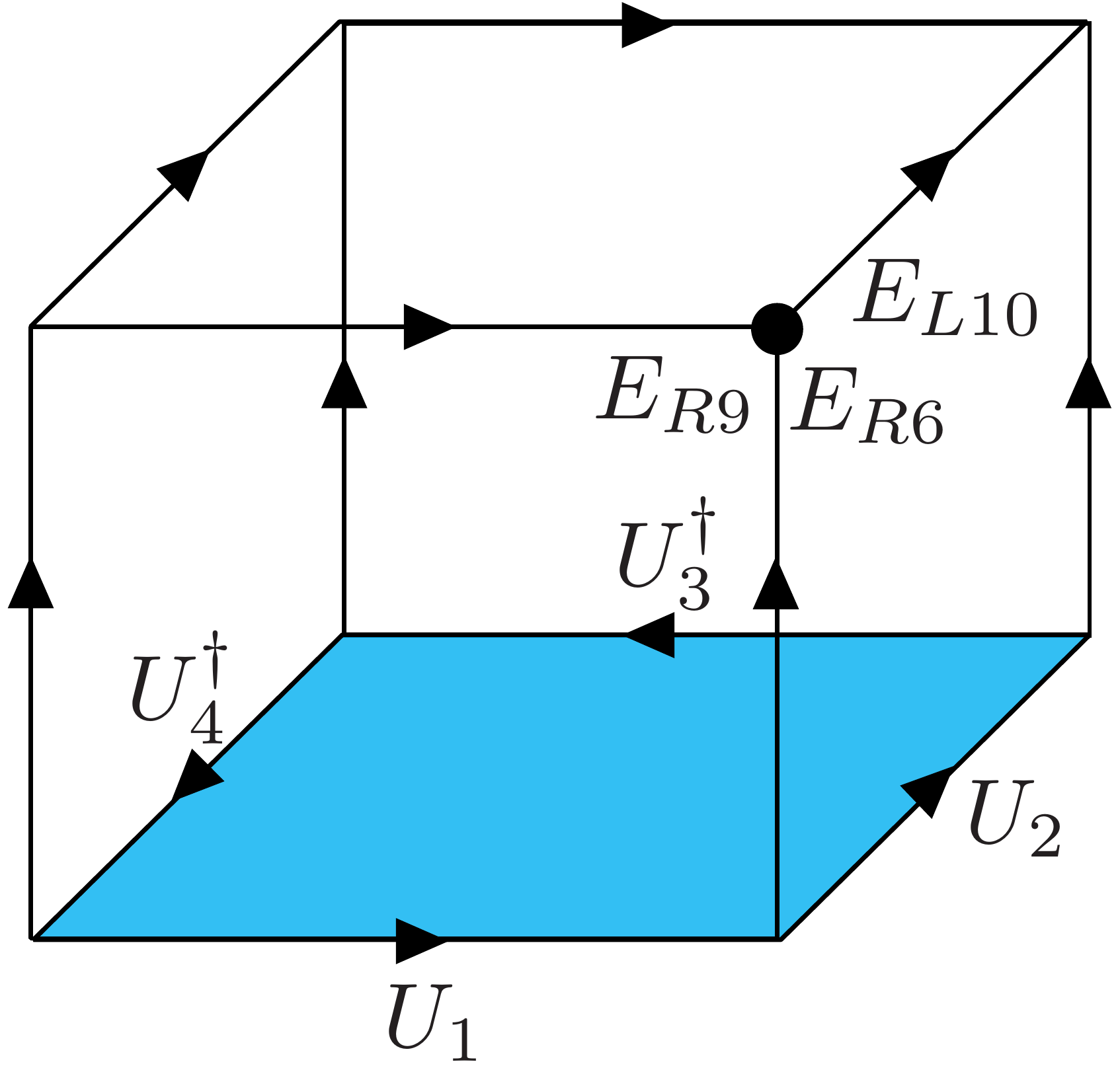}
   \caption{Schematic representation of the lattice SU(2) Yang-Mills theory. The gauge field $U_\mu(\bm x)$ is defined on a link.
   The chromo-electric fields $E_L$ and $E_R$ are defined on the ends of a link, and satisfy the Gauss law constraint on the vertices, e.g., $E_{R6}+E_{R9}+E_{L10}=0$ at the vertex indicated by the black dot.
   The minimal Wilson loop operator is defined as the product of the link operators on the edges of a unit plaquette colored in the figure. }
  \label{fig1}
  \end{figure}

\section{Hamiltonian formulation}\label{sec:Hamitonian formalism}
\subsection{formulation}
We review the Hamiltonian formulation of the lattice SU($2$) Yang-Mills theory, which is often referred to as the Kogut-Susskind Hamiltonian formulation~\cite{Kogut:1974ag}.
We consider a cubic lattice (See Fig.~\ref{fig1}).
The gauge fields $U_\mu(\bm x)$ are defined on a link emanating from a site $\bm x$ and terminating at $\bm x+\hat{e}_\mu$, with $\hat{e}_{\mu=x,y,z}$ being the unit vector along the $\mu$ direction.
$U_\mu(\bm x)$ is a $2\times2$ matrix-valued operator, and we can apply the local SU($2$) transformation to $U_\mu(\bm x)$ from left-/right-hand side.
Using the generators $E^{a}_L(\bm x,\mu)$, and $E^a_R(\bm x,\mu)$ [$a=x,y,z$], which are nothing but the left- or right-chromoelectric fields defined on the ends of a link, the SU(2) algebras are represented as
\begin{align}
  [E^{a}_L(\bm x,\mu),U_\nu(\bm y)] &=-\frac{1}{2}\sigma^aU_\mu(\bm x)\delta_{\mu\nu}\delta_{\bm x,\bm y} , \label{eq:[E_L,U]}
  \\
  [E^{a}_R(\bm x,\mu),U_\nu(\bm y)] &=U_\mu(\bm x) \frac{1}{2}\sigma^a\delta_{\mu\nu}\delta_{\bm x,\bm y},\\
[E^{a}_{L}(\bm x,\mu),E^{b}_{L}(\bm x,\nu)] &=\ri\epsilon^{abc}E^{c}_{L}(\bm x,\mu)\delta_{\mu\nu}\delta_{\bm x,\bm y},     \\
[E^{a}_{R}(\bm x,\mu),E^{b}_{R}(\bm x,\nu)] &=\ri\epsilon^{abc}E^{c}_{R}(\bm x,\mu)\delta_{\mu\nu}\delta_{\bm x,\bm y},     \label{eq:[E_R,E_R]}
\end{align}
where $\sigma^{a=x,y,z}$ are the Pauli matrices, and other commutation relations vanish.
$\epsilon^{abc}$ is the Levi-Civita symbol with $\epsilon^{xyz}=1$.
In the Kogut-Suskind Hamiltonian formulation, the generators are not independent but related through the constraint
\bea
\begin{split}
 \sum_aE^{a}_L(\bm x,\mu)E^{a}_L(\bm x,\mu)
 &=\sum_a E^{a}_R(\bm x,\mu)E^{a}_R(\bm x,\mu)\\
 &=:E^2(\bm x,\mu).
\end{split}
\label{eq:constraint}
\eea
The Hamiltonian is given as the sum of electric and magnetic parts, $\cH=\cH_E+\cH_B$, with
\begin{align}
  &\cH_E=\sum_{\bm x,\mu}\frac{1}{2}E^2(\bm x,\mu),
  \label{eq:electric}
  \\
  &\cH_B=
   -\frac{K}{2}\sum_{p\in P}\tr[U_{\mu}(\bm x)U_{\nu}(\bm x+\hat{e}_\mu)U_{\mu}^{\dagger}(\bm x+\hat{e}_\nu)U_{\nu}^{\dagger}(\bm x)]\notag\\
   &\quad\qquad +({\rm h.c.}) ,
  \label{eq:magnetic}    
\end{align}
where $P$ is the set of plaquettes, and 
``h.c.'' represents the Hermitian conjugate. $K$ is the coupling constant, which is inversely proportional to the square of the gauge coupling $g$. Therefore, we use the words, strong and weak coupling, for small and large $K$, respectively.
$\cH_E$ is the electric part of the Hamiltonian, which has the same form as continuum theory.
$\cH_B$ is the lattice version of the magnetic part of the Hamiltonian; it involves a nonlocal but gauge-invariant operator of $U_{\mu}(x)$, which is the famous Wilson loop operator (see Fig.~\ref{fig1}).
The Schr{\"o}dinger equation with the Hamiltonian defines the dynamics on the physical Hilbert space $|\Psi\rangle$ that satisfies the Gauss law constraints: 
\bea
\sum_\mu\left( E^a_L(\bm x,\mu)+E^{a}_R(\bm x-\hat{e}_\mu,\mu) \right)|\Psi\rangle=0 .
\label{eq:gauss}
\eea
The Gauss law constraints simply state that the total electric field at a site $\bm x$ must vanish.

For numerical implementation, we rewrite the  Kogut-Susskind Hamiltonian using the so-called Schwinger bosons~\cite{Mathur:2004kr}.
In SU($2$), the Hamiltonian of the chromoelectric fields~\eqref{eq:electric} is the same as that of the quantum rotor.
Therefore, the electric field operator can be understood as the angular momentum operator, and represented by using the creation and annihilation operators of the spin doublet bosons (Schwinger bosons) as
\begin{align}
  E^a_{L}(\bm x,\mu) &=a_{i}^\dagger(\bm x,\mu)\frac{1}{2}{\sigma}^a_{ij} a_j(\bm x,\mu), 
  \label{eq:E_left}
  \\
  E^a_{R}(\bm x,\mu)&=b_{i}^\dagger(\bm x,\mu)\frac{1}{2}\sigma^a_{ij} b_j(\bm x,\mu) ,
  \label{eq:E_right}
\end{align}
where $a_{i=\up,\down}$,  $a^\dagger_{i=\up,\down}$ ($b_{i=\up,\down}$, and $b^\dagger_{i=\up,\down}$) are the annihilation and creation operators of the Schwinger bosons, which are defined on the left (right) end of a link.
In terms of the Schwinger bosons, the constraint~\eqref{eq:constraint} implies
\be
N_L(\bm x,\mu)|\Psi\rangle=N_R(\bm x,\mu) |\Psi\rangle,
\ee
where $N_L=\sum_iN_{Li}=\sum_i a_{i}^\dagger a_i$, and $N_R=\sum_iN_{Ri}=\sum_i b_{i}^\dagger b_i$ are the number operators of the Schwinger bosons.
Therefore, the total number of Schwinger bosons living on the edges of a link must be the same.
Using the Schwinger boson representation, the electric part of the Hamiltonian is written only by $N_L$ or $N_R$. 

Next, we consider the magnetic part of the Hamiltonian.
It is known that the link operator can be written using the Schwinger bosons as
$U=U_L(a)U_R(b)$ with
 \begin{align}
  U_L(a) &=
  \frac{1}{\sqrt{N_L+1}}
  \begin{pmatrix}
  a_\down^\dagger & a_\up \\
  -a_\up^\dagger & a_\down
  \end{pmatrix} ,
  \label{eq:linkL}
   \\
  U_R(b) &=
  \begin{pmatrix}
  b_\up^\dagger & b_\down^\dagger \\
  -b_\down & b_\up
  \end{pmatrix}
  \frac{1}{\sqrt{N_R+1}} .
  \label{eq:linkR}     
 \end{align}
Using the commutation relations between creation and annihilation operators , Eqs.~\eqref{eq:E_left},~\eqref{eq:E_right}, \eqref{eq:linkL} and \eqref{eq:linkR}
reproduce those between $E_{L,R}$, and $U$ in Eqs.~\eqref{eq:[E_L,U]}-\eqref{eq:[E_R,E_R]}.

We can label the Hilbert space of the gauge theory by the number of eigenvectors of the harmonic oscillators.
This enables us to understand the complex wave function of the gauge theory from the simple picture of the occupation dynamics of bosons. Furthermore, by truncating the max occupation number of the Schwinger bosons, we can obtain the finite-dimensional Hilbert space with manifestly keeping the gauge symmetry.

Two remarks are in order: (I) Truncating the Schwinger boson occupation numbers at a certain value is equivalent to representing the Kogut-Susskind theory with the $j$-dimensional irreducible representation of SU($2$) up to $j_{\rm max}=(N_L+N_R)/2$.
The limit $j_{\rm max}\rightarrow\infty$ may recover Wilson's formulation of lattice gauge theories. 
(II) The magnetic part of the Hamiltonian~\eqref{eq:magnetic} does change the number of the Schwinger bosons on each link with satisfying the constraint~\eqref{eq:constraint}, while the electric part of the Hamiltonian~\eqref{eq:electric} just counts their numbers.
Therefore, they can be understood as the kinetic and interaction terms of the Schwinger bosons.
Without magnetic interactions ($K=0$) corresponding to the strong coupling limit, the gauge theory is reduced to free harmonic oscillators.
As $K$ increases, the fluctuations by the magnetic Hamiltonian become relevant, and then the gauge theory becomes strongly correlated.
In what follows, we study such strongly correlated dynamics of the Yang-Mills theory by quenching the magnetic Hamiltonian and solving the time-dependent Schr{\"o}dinger equation after the quench. We note that larger $K$ demands larger $j_{\rm max}$ for full quantitative analysis since the effect of the truncation of the Hilbert space becomes more relevant  as the typical occupation number increases.

\begin{figure}[h]
  \centering
  \includegraphics[width=.25\textwidth]{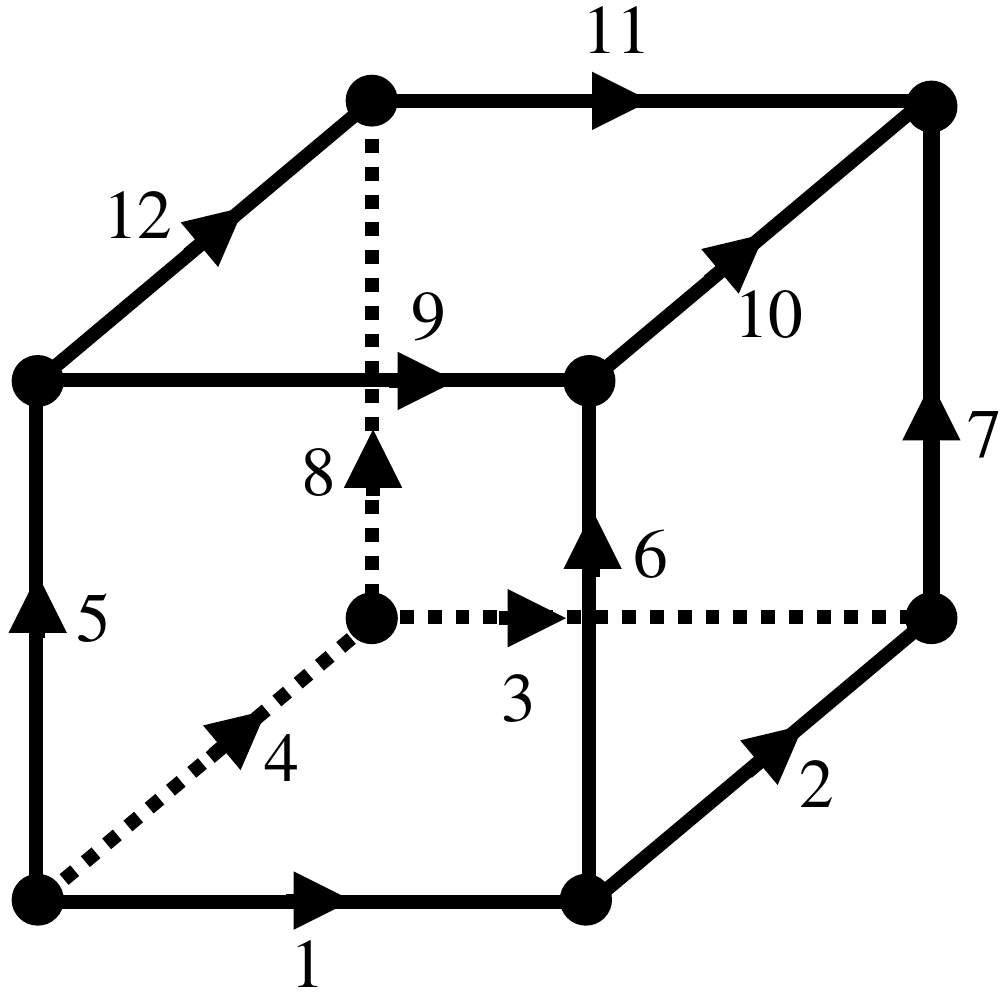}
  \caption{Labels of links on the single cubic lattice model.}
  \label{fig2}
 \end{figure}
 
\subsection{Construction of physical states}
 Here, we explicitly construct the physical states on a single-cubic lattice by solving the Gauss law and $U(1)$ constraints (see Refs.~\cite{Mathur:2004kr,Mathur:2005fb,Mathur:2007nu,Anishetty:2014tta,Raychowdhury:2018tfj,Raychowdhury:2019iki} for more general and other constructions of the Hilbert space). For convenience, we label links by integers $L\coloneqq\{1,2,\cdots,12\}$ as shown in Fig.~\ref{fig2}. We express the vertices and the plaquettes by ordered triples, $V\coloneqq\{(1,2,6)$,$(2,3,7)$,$(3,4,8)$,$(1,4,5)$,$(6,9,10)$,$(7,10,11)$,
 $(8,11,12)$,$(5,9,12)\}$,
 and by ordered quadruples, $P=
 \{(1,2,3,4)$,$(1,6,9,5)$,$(2,7,10,6)$,$(3,7,11,8)$,$(4,8,12,5)$,
 $(9,10,11,12)\}$, respectively.
 
 The Gauss law constraint~\eqref{eq:gauss} needs to be satisfied at each vertex. Let us focus on the vertex $(1,2,6)\in V$. The local state is expressed by 
 $|N_{1\uparrow} \, N_{1\downarrow}\rangle|N_{2\uparrow} \, N_{2\downarrow}\rangle|N_{6\uparrow} \, N_{6\downarrow}\rangle=|j_1 \, m_1\rangle|j_2 \, m_2\rangle|j_6 \, m_6\rangle$
 with $j_a=(N_{a\uparrow}+N_{a\downarrow})/2$ and 
 $m_a=(N_{a\uparrow}-N_{a\downarrow})/2$ ($a=1,2,6$).
 The spin basis is useful for solving the Gauss law constraint, while
 the number basis is useful for calculating the matrix elements. We use both representations in the following.
 The Gauss law constraint implies that the local state is the spin singlet. Since a state of the vertex is a composition of three spin states, we can express the singlet state by using the Wigner $3$-$j$ symbols 
 $\begin{psmallmatrix}
  j_1 & j_2 & j_6\\
  m_1 & m_2 & m_6
 \end{psmallmatrix}$ as
 \begin{equation}
  \begin{split}
 |j_1,j_2,j_6\rangle &\coloneqq  \sum_{m_1=-j_1}^{j_1}\sum_{m_2=-j_2}^{j_2}\sum_{m_6=-j_6}^{j_6} 
 \begin{pmatrix}
  j_1 & j_2 & j_6\\
  m_1 & m_2 & m_6
 \end{pmatrix}\\
&\quad \times|j_1,m_1\rangle|j_2,m_2\rangle|j_6,m_6\rangle,
\end{split}
 \end{equation}
 which satisfies the standard normalization condition $\langle j_i,j_j,j_k|j_{l},j_{m},j_{n}\rangle=\delta_{il}\delta_{jm}\delta_{kn}$.
 We note that $|j_1,j_2,j_6\rangle$ is nonvanishing only if the triangle conditions $|j_1-j_2|\leq j_6\leq j_1+j_2$ and $j_1+j_2+j_6\in \mathbb{Z}$ are satisfied.
 We also note that the Wigner $3$-$j$ symbols have the symmetry properties under permutations:
 $\begin{psmallmatrix}
  j_1 & j_2 & j_6\\
  m_1 & m_2 & m_6
 \end{psmallmatrix}
 =
 \begin{psmallmatrix}
  j_2 & j_6 & j_1\\
  m_2 & m_6 & m_1
 \end{psmallmatrix}
 =
 \begin{psmallmatrix}
  j_6 & j_1 & j_2\\
  m_6 & m_1 & m_2
 \end{psmallmatrix}
 =
 (-1)^{j_1+j_2+j_6}
 \begin{psmallmatrix}
  j_2 & j_1 & j_6\\
  m_2 & m_1 & m_6
 \end{psmallmatrix}
 =
 (-1)^{j_1+j_2+j_6}
 \begin{psmallmatrix}
  j_1 & j_6 & j_2\\
  m_1 & m_6 & m_2
 \end{psmallmatrix}
 =
 (-1)^{j_1+j_2+j_6}
 \begin{psmallmatrix}
  j_6 & j_2 & j_1\\
  m_6 & m_2 & m_1
 \end{psmallmatrix}$.
 To respect the symmetry properties, we introduce the permutated states defined as
 \begin{equation}
  |j_{i},j_j,j_k\rangle \coloneqq (\mathrm{sgn}(\sigma))^{j_1+j_2+j_6}|j_{1},j_2,j_6\rangle,
  \label{eq:permutation}
 \end{equation}
 where $\mathrm{sgn}(\sigma)$ is the sign of permutation $\sigma=\begin{psmallmatrix}1&2&6\\ i&j&k  \end{psmallmatrix}$ (Not to be confused with $3$-$j$ symbol).
 On the other hand, the $U(1)$ constraint $N_L(\bm{x},\mu)|\Psi\rangle=  N_R(\bm{x},\mu)|\Psi\rangle$ implies the vertices connecting to a link shear the same spin $j$.
 For example, for the vertices $(1,4,5)$ and $(1,2,6)$ connecting to the link $1$, the local states are expressed as $|j_1',j_4,j_5\rangle$ and $|j_1,j_2,j_6\rangle$, respectively.
 The $U(1)$ constraint means $j_1'=j_1$.
 Eventually, we can express a physical state by using the states of spins on links, $\bm{j}=(j_1,\cdots j_{12})$ as
 \begin{equation}
  \begin{split}
  |\bm{j}\rangle
  &\coloneqq \prod_{(i,j,k)\in V}|j_i,j_j,j_k\rangle\\
  &=
     |j_1,j_2,j_6\rangle  |j_2,j_3,j_7\rangle  |j_3,j_4,j_8\rangle |j_1,j_4,j_5\rangle
      |j_{6},j_9,j_{10}\rangle \\
      &\quad\times |j_{7},j_{10},j_{11}\rangle|j_{8},j_{11},j_{12}\rangle|j_{5},j_{9},j_{12}\rangle  .
  \label{eq:physical state}
  \end{split}
 \end{equation}
 Since $j_i$ has no upper bound, the dimension of the physical Hilbert space is infinite.
 In numerical simulations, we truncate the spin's maximum value, $j_\text{max}$, to make the dimension of the Hilbert space finite.
 We show the $j_\text{max}$ dependence of the dimension of the physical Hilbert space in Table~\ref{tab:dimension}.
 
 \begin{table*}[htb]
  \begin{tabular}{l|ccccccccc}
    $j_\text{max}$ & 0 &  1/2  &  1  &  3/2 & 2 & 5/2 & 3 & 7/2   \\
    \hline
     $d$ & 1 & 32 & 1013 & 14,879 & 148,678 & 1,007,699 & 5,410,350 & 23,403,554   \\
         $j_\text{max}$ & 4 &  9/2 & 5 &  11/2 & 6 & 13/2 & 7 & 15/2   \\
    \hline
     $d$ & 87,426,119 & 285,115,818 & 841,734,227 & 2,264,663,617  & 5,671,695,596  & 13,279,002,317  & 29,457,092,444 & 62,092,681,444      
  \end{tabular}
  \caption{$j_\text{max}$ dependence of the dimension of the physical Hilbert space $d$.\label{tab:dimension}}
 \end{table*}
  
 \subsection{Matrix element of Hamiltonian}
 Let us evaluate the matrix element of the Hamiltonian $H=H_E+H_M$.
 Because $|\bm{j}\rangle$ is an eigenstate of $H_E$, the matrix element of $H_E$ is just the sum of eigenvalues of spins:
 \begin{equation}
  \langle \bm{j}'|H_E|\bm{j}\rangle = \sum_{i\in L}\frac{j_i(j_i+1)}{2}\delta_{\bm{j}',\bm{j} }.
 \end{equation}
 For the magnetic part, more calculations are involved. Since $H_M$ consists of the sum of plaquettes, let us, first, focus on the single plaquette $ \mathrm{tr}(U_{1}U_{2}U_{3}^\dag U_{4}^\dag)$. Noting $U_{i}=U_L(a_i)U_R(b_i)$,
 we can write $\mathrm{tr}(U_{1}U_{2}U_{3}^\dag U_{4}^\dag)$ as
 \begin{equation}
  \begin{split}
  \mathrm{tr}(U_{1}U_{2}U_{3}^\dag U_{4}^\dag)&=
  \mathrm{tr}\bigl([U_R(b_1)U_L(a_2)][U_R(b_2)U_R^\dag(b_3)]\\
  &\qquad\times[U_L^\dag(a_3)U_R^\dag(b_4)][U_L^\dag(a_4)U_L(a_1)]\bigr).
  \end{split}
 \end{equation}
 Here, we employed the cyclic property of the trace.
 It is useful to express the matrices on each vertex as
 \begin{align}
  U_R(b_1)U_L(a_2)
 &=\begin{pmatrix}
 \mathcal{L}^{++}(b_1,a_2) & \mathcal{L}^{+-}(b_1,a_2)\\
 \mathcal{L}^{-+}(b_1,a_2) & \mathcal{L}^{--}(b_1,a_2)
 \end{pmatrix},\\
 \quad  U_R(b_2) U_R^\dag(b_3)
& =
 \begin{pmatrix}
 \mathcal{L}^{+-}(b_2,b_3) & -\mathcal{L}^{++}(b_2,b_3)\\
 \mathcal{L}^{--}(b_2,b_3) & -\mathcal{L}^{-+}(b_2,b_3)
 \end{pmatrix},\\
 U_L^\dag(a_3) U_R^\dag(b_4)
 &=
 \begin{pmatrix}
 -\mathcal{L}^{--}(a_3,b_4) & \mathcal{L}^{-+}(a_3,b_4)\\
 \mathcal{L}^{+-}(a_3,b_4) & -\mathcal{L}^{++}(a_3,b_4)
 \end{pmatrix},\\
 U_L^\dag(a_4) U_L(a_1)
& =
 \begin{pmatrix}
 -\mathcal{L}^{-+}(a_4,a_1) & -\mathcal{L}^{--}(a_4,a_1)\\
 \mathcal{L}^{++}(a_4,a_1) & \mathcal{L}^{+-}(a_4,a_1)
 \end{pmatrix},
 \end{align}
 where we define
 \begin{align}
  \mathcal{L}^{++}(b,a)&=
  \frac{1}{\sqrt{N_L+1}}
  (b_{\uparrow}^\dag a_{\downarrow}^\dag -b_{\downarrow}^\dag a_{\uparrow}^\dag)
  \frac{1}{\sqrt{N_R+1}},
  \\
  \mathcal{L}^{-+}(b,a)&=
  \frac{1}{\sqrt{N_L+1}}
  (-b_\downarrow a_{\downarrow}^\dag-b_{\uparrow} a_{\uparrow}^\dag)
  \frac{1}{\sqrt{N_R+1}},
  \\
 \mathcal{L}^{+-}(b,a)&=
 \frac{1}{\sqrt{N_L+1}}
 (b_{\uparrow}^\dag a_{\uparrow}+b_\downarrow^\dag a_{\downarrow})
 \frac{1}{\sqrt{N_R+1}},
 \\
 \mathcal{L}^{--}(b,a)&=
 \frac{1}{\sqrt{N_L+1}}
 (-b_{\downarrow} a_{\uparrow}+b_{\uparrow} a_{\downarrow})
 \frac{1}{\sqrt{N_R+1}}.
 \end{align}
 This expression enables us to express the plaquette as the sum of $\mathcal{L}$'s,
 \begin{equation}
  \begin{split}
  \mathrm{tr}(U_{1}U_{2}U_{3}^\dag U_{4}^\dag)&=\sum_{s_1,s_2,s_3,s_4=\pm1}
  {\mathcal{L}}^{s_1s_2}(b_1,a_2){\mathcal{L}}^{s_2s_3}(b_2,b_3)\\ 
 &\qquad\qquad\qquad\times{\mathcal{L}}^{s_3s_4}(a_3,b_4) {\mathcal{L}}^{s_4s_1}(a_4,a_1).
  \end{split}
 \end{equation}
 Each ${\mathcal{L}}^{s's}(b,a)$ locally acts on the Hilbert space, so that it is enough to consider the action of ${\mathcal{L}}^{s's}(b,a)$ on a single vertex.
 Let $| j_1,j_2,j_{6}\rangle$ be a local state on the vertex $(1,2,6)\in V$.
 Since ${\mathcal{L}}^{s's}(b,a)$ is written by the creation and annihilation operators,
 it is easy to calculate the action of ${\mathcal{L}}^{s's}(b,a)$.
 Remembering $|j,m\rangle=|N_{\uparrow},N_{\downarrow}\rangle$ with $j=(N_{\uparrow}+N_{\downarrow})/2$ and $m=(N_{\uparrow}-N_{\downarrow})/2$, we can calculate
 $a_{\uparrow}^{\dag}|j,m\rangle$ as
 \begin{equation}
  \begin{split}
  a_{\uparrow}^{\dag}|j,m\rangle
  &=a_{\uparrow}^{\dag}|N_{\uparrow},N_{\downarrow}\rangle\\
  &=\sqrt{N_\uparrow+1}|N_{\uparrow}+1,N_{\downarrow}\rangle\\
  &=\sqrt{j+m+1}|j+\frac{1}{2},m+\frac{1}{2}\rangle.
  \end{split}
 \end{equation}
 Similarly, we obtain
 \begin{align}
  a_{\uparrow}|j,m\rangle
  &=\sqrt{j+m}|j-\frac{1}{2},m-\frac{1}{2}\rangle,\\
  a_{\downarrow}^{\dag}|j,m\rangle
  &=\sqrt{j-m+1}|j+\frac{1}{2},m-\frac{1}{2}\rangle,\\
  a_{\downarrow}|j,m\rangle
  &=\sqrt{j-m}|j-\frac{1}{2},m+\frac{1}{2}\rangle.
 \end{align}
 Using these relations, we find
 \begin{equation}
  \begin{split}
  &\mathcal{L}^{s_1s_2}(b_1,a_2)
    |j_1,j_2,j_6\rangle\\
     &=  \sum_{m_1=-j_1}^{j_1}\sum_{m_2=-j_2}^{j_2}\sum_{m_6=-j_6}^{j_6} 
    \begin{pmatrix}
      j_1 & j_2 & j_6\\
      m_1 & m_2 & m_6
    \end{pmatrix}\\
    &\quad\times
    \Bigl(
      s_1\sqrt{\frac{(j_1+s_1m_1+\frac{1+s_1}{2})(j_2-s_2m_2+\frac{1+s_2}{2})}{(2j_1+1)(2j_2+1+s_2)}}\\
      &\quad\quad\times|j_1+\frac{s_1}{2},m_1+\frac{1}{2}\rangle|j_2+\frac{s_2}{2},m_2-\frac{1}{2}\rangle\\
      &\quad-s_2\sqrt{\frac{(j_1-s_1m_1+\frac{1+s_1}{2})(j_2+s_2m_2+\frac{1+s_2}{2})}{(2j_1+1)(2j_2+1+s_2)}}\\
      &\quad\quad\times|j_1+\frac{s_1}{2},m_1-\frac{1}{2}\rangle|j_2+\frac{s_2}{2},m_2+\frac{1}{2}\rangle
    \Bigr)|j_6,m_6\rangle.
  \end{split}
  \label{eq:L}
 \end{equation}
 Since $\mathcal{L}^{s_1s_2}(b_1,a_2)$ commutes with the Gauss law constraint, the right-hand side of Eq.~\eqref{eq:L} must be proportional to $| j_1+{s_1}/{2},j_2+{s_2}/{2},j_{6}\rangle$, i.e.,
 \begin{equation}
  \begin{split}
    &\mathcal{L}^{s_1s_2}(b_1,a_2)| j_1,j_2,j_{6}\rangle\\
    &=\lambda_{s_1s_2}(j_1,j_2,j_6)| j_1+\frac{s_1}{2},j_2+\frac{s_2}{2},j_{6}\rangle,
  \end{split}
 \end{equation}
 is satisfied.
By comparing the wave functions of $\mathcal{L}^{s_1s_2}(b_1,a_2)| j_1,j_2,j_{6}\rangle$ and $| j_1+{s_1}/{2},j_2+{s_2}/{2},j_{6}\rangle$, we obtain
 \begin{align}
  \lambda_{++}(j_1,j_2,j_6)&=\sqrt{\frac{(2+j_6+j_1+j_2)(1-j_6+j_1+j_2)}{(2j_1+1)(2j_2+2)}},\\
  \lambda_{--}(j_1,j_2,j_6)
 &=\sqrt{ \frac{(j_6-j_1-j_2)(-1-j_1-j_2-j_6)}{(2j_1+1)(2j_2)} },\\
 \lambda_{+-}(j_1,j_2,j_6)
 &=-\sqrt{\frac{(1+j_6+j_1-j_2)(j_6-j_1+j_2)}{(2j_1+1)(2j_2)}},\\
 \lambda_{-+}(j_1,j_2,j_6)
 &= \sqrt{\frac{(1+j_6-j_1+j_2)(j_6+j_1-j_2)}{(2j_1+1)(2j_2+2)} }.
 \end{align}
Similarly, we also obtain
 \begin{align}
  &\mathcal{L}^{s_2s_3}(b_2,b_3)|j_2,j_3,j_7\rangle\notag\\
  &=\lambda_{s_2s_3}(j_2,j_3,j_7)|j_2+\frac{s_2}{2},j_3+\frac{s_3}{2},j_7\rangle\\
  &\mathcal{L}^{s_3s_4}(a_3,b_4)|j_3,j_4,j_8\rangle\notag\\
  &=\lambda_{s_3s_4}(j_3,j_4,j_8)|j_3+\frac{s_3}{2},j_4+\frac{s_4}{2},j_8\rangle.
 \end{align}
 In contrast, we have to be careful for the calculation of $\mathcal{L}^{s_4s_1}(a_4,a_1)|j_1,j_4,j_5\rangle$ because the ordering $(4,1)$ in $\mathcal{L}^{s_4s_1}(a_4,a_1)$ is opposite to $(1,4,5)$ in $|j_1,j_4,j_5\rangle$ .
 This can be done by using the permutation property~\eqref{eq:permutation} as
 \begin{equation}
  \begin{split}
 &\mathcal{L}^{s_4s_1}(a_4,a_1)|j_1,j_4,j_5\rangle\\
 &=\mathcal{L}^{s_4s_1}(a_4,a_1)(-1)^{j_1+j_4+j_5}|j_4,j_1,j_5\rangle\\
 &=\lambda_{s_4s_1}(j_4,j_1,j_5)(-1)^{j_1+j_4+j_5}|j_4+\frac{s_4}{2},j_1+\frac{s_1}{2},j_5\rangle\\
 &=(-1)^{\frac{s_1+s_4}{2}}\lambda_{s_4s_1}(j_4,j_1,j_5)|j_1+\frac{s_1}{2},j_4+\frac{s_4}{2},j_5\rangle.
  \end{split}
  \label{eq:permutation}
 \end{equation}
 Here we used the fact that $2(j_1+j_4+j_6)$ is an even integer in the last line.
 We obtained the phase factor $(-1)^{({s_1+s_4})/{2}}$ in addition to $\lambda_{s_4s_1}(j_4,j_1,j_5)$. We note that this phase factor comes from the signature of the permutation $(1,4,5)\to (4,1,5)$.
 From these results, we get
 \begin{equation}
  \begin{split}
  &\mathrm{tr}(U_{1}U_{2}U_{3}^\dag U_{4}^\dag)| j_1,j_2,j_{6}\rangle| j_2,j_3,j_{7}\rangle| j_3,j_4,j_{8}\rangle| j_4,j_1,j_{5}\rangle\\
  &=\sum_{s_1,s_2,s_3,s_4=\pm 1}
  {\lambda}_{s_1s_2}(j_1,j_2,j_{6})
  {\lambda}_{s_2s_3}(j_2,j_3,j_{7})\\
  &\quad\times{\lambda}_{s_3s_4}(j_3,j_4,j_{8})
  {\lambda}_{s_4s_1}(j_4,j_1,j_{5}) 
  (-1)^{\frac{s_4+s_1}{2}}
  \\
  &\quad\times| j_1+\frac{s_1}{2},j_2+\frac{s_2}{2},j_{6}\rangle
  | j_2+\frac{s_2}{2},j_3+\frac{s_3}{2},j_{7}\rangle\\
  &\quad\times| j_3+\frac{s_3}{2},j_4+\frac{s_4}{2},j_{8}\rangle
  |j_1+\frac{s_1}{2}, j_4+\frac{s_4}{2},j_{5}\rangle.
  \end{split}
 \end{equation}
 In the same way, we can show $\mathrm{tr}(U_{1}U_{2}U_{3}^\dag U_{4}^\dag)|\bm{j}\rangle=\mathrm{tr}(U_{4}U_{3}U_{2}^\dag U_{1}^\dag)|\bm{j}\rangle$.

 In order to evaluate other plaquettes, we need the formula of the phase factor.
 For a given plaquette $(i,j,k,l)=p\in P$, we can define the set of vertices: $V_p=\{
  (i,j,c_{ij}), (j,k,c_{jk}),(j,k,c_{jk}),(k,l,c_{kl}),(l,i,c_{li})  \}$, where $c_{mn}$ is the link that shares the vertex with the links $m$ and $n$.
  For $(i,j,c_{ij})\in V_p$, there exists the corresponding vertex $(a,b,c)\in V$ that differs only the ordering from $(i,j,c_{ij})$.
  Equation~\eqref{eq:permutation} implies that the phase factor comes from the signature of permutations, so that we define the signature of $(i,j,c_{ij})\in V_p$ as 
  \begin{equation}
    \begin{split}
    &\mathrm{sgn}(i,j)\\
     &\coloneqq
    \begin{cases}
      +1 & \text{if $(i,j,c_{ij})$ is the even permuation of $(a,b,c)$}\\
      -1 & \text{if $(i,j,c_{ij})$ is the odd permuation of $(a,b,c)$}\\
    \end{cases}
  \end{split}.
  \end{equation}
  The phase factor of the action $\mathcal{L}^{s_is_j}$ is given as $(\mathrm{sgn}(i,j))^{(s_i+s_j)/2}$.
  For example, for the plaquette $p=(1,6,9,5)$, we have $V_{p}=\{(1,6,2),(6,9,10),(9,5,12),(5,1,4) \}$, and 
  the corresponding vertex of $(1,6,2)\in V_p$ is $(1,2,6)\in V$. Then, $c_{16}$ and $\mathrm{sgn}(1,6)$ are 
  $c_{16}=2$, and $\mathrm{sgn}(1,6)=-1$.
 
  Using these definitions, we find the matrix element of the magnetic part as
 \begin{equation}
  \begin{split}
 \langle \bm{j}'|H_M|\bm{j}\rangle&= -K\sum_{p\in P}\sum_{s_l=\pm1,l\in p}\prod_{(i,j,c_{ij})\in V_p}\\
&\quad\times (\mathrm{sgn}(i,j) )^{\frac{s_i+s_j}{2}}
 \lambda_{s_i,s_j}(j_i,j_j,j_{c_{ij}}) \delta_{\bm{j}',\bm{j}+\frac{\bm{s}_p}{2}},
  \end{split}
 \end{equation}
 where $\bm{s}_p$ is a vector whose components are defined as
 \begin{equation}
 (\bm{s}_p)_i
 =\begin{cases}
 s_i\in\{1,-1\}  &\text{if}\quad i\in p\\
 0  &\text{else}
 \end{cases}.
 \end{equation}
 In summary, the physical state is given in Eq.~\eqref{eq:physical state}, and the matrix element of the Hamiltonian is expressed as
 \begin{equation}
  \begin{split}
  \langle \bm{j}'|H|\bm{j}\rangle&= \sum_{i\in L}\frac{j_i(j_i+1)}{2}\delta_{\bm{j}',\bm{j} }
  -K\sum_{p\in P}\sum_{s_l=\pm1,l\in p}\prod_{(i,j,c_{ij})\in V_p}\\
  &\times(\mathrm{sgn}(i,j) )^{\frac{s_i+s_j}{2}}
  \lambda_{s_i,s_j}(j_i,j_j,j_{c_{ij}}) \delta_{\bm{j}',\bm{j}+\frac{\bm{s}_p}{2}}.
  \end{split}
 \end{equation}

\section{Real-time simulation}\label{sec:Real time simulation}
\subsection{Interaction quench}
We consider the single cubic lattice with open boundary conditions, shown in Fig.~\ref{fig1}.
We truncate the Schwinger boson occupation number at $j_\text{max}=(N_R+N_L)/2$.
For example, if we consider the lowest truncation $j_\text{max}=1/2$, the dimension of the local Hilbert space is $5$.
In the number basis $|N_{L\up} N_{L\down}\rangle |N_{R\up} N_{R\down}\rangle$, these are explicitly given as 
$|00\rangle|00\rangle, \; |10\rangle|10\rangle, \; |10\rangle|01\rangle, \; |01\rangle|10\rangle, \; |01\rangle|01\rangle$.
Therefore, the dimension of the full Hilbert space is $5^{12}\sim$ $0.2$ billion.
The full Hilbert space is so large that we cannot manage it in numerical simulations except the lowest $j_\text{max}$.
However, the majority of the Hilbert space represents the redundancy associated with the gauge symmetry, and we need only the subspace (physical Hilbert space) obtained by solving the Gauss law constraints~\eqref{eq:gauss}.
This process significantly reduces the dimension of the Hilbert space, e.g., from $5^{12}$ to $32$ for $j_\text{max}=1/2$.
By explicitly solving the Gauss law constraints, we can do numerical simulations with larger $j_\text{max}$.

We numerically solve the time-dependent Schr{\"o}dinger equation, $\ri\partial_t |\Psi(t)\rangle=H|\Psi(t)\rangle$, in the physics Hilbert space and Hamiltonian constructed in the previous section.
As an initial state, we choose the Fock vacuum defined by $a_i(\bm x,\mu)| \Psi(0)\rangle=b_i(\bm x,\mu)| \Psi(0)\rangle=0$.
The Fock vacuum is the eigenstate of the electric Hamiltonian~\eqref{eq:electric}, that is, the ground state of the Hamiltonian at the strong coupling limit $K=0$.
We study the real-time dynamics after the magnetic Hamiltonian is switched on at $t=0$.
We solved  the time-dependent Schr{\"o}dinger equation based on the leap-frog type discretization.
We decompose the time-dependent Schr{\"o}dinger equation into two real-valued equations:
\bea
\partial_t {\rm Re}[|\Psi(t)\rangle] &=&\;\;\; H {\rm Im}[|\Psi(t)\rangle] , 
\\
\partial_t {\rm Im}[|\Psi(t)\rangle] &=& -H{\rm Re}[|\Psi(t)\rangle] .
\eea
We regard the real and imaginary parts as ``position" and ``velocity" and apply the leap-frog integrator.
This method is applicable only when the Hamiltonian is real-valued in some basis\footnote{It was found to be less efficient in our case, but for a generic Hamiltonian, we can solve the unitary evolution $|\Psi(t)\rangle=e^{-iHt}|\Psi(0)\rangle$ based on the Krylov subspace method.}.
The numerical resources needed to obtain the following results, e.g., with $j_{\rm max}=4$ ($d=87,426,119$) are 262 TFlops*hr for each $K$.

  \begin{figure}[t]
    \centering
    \includegraphics[width=.41\textwidth]{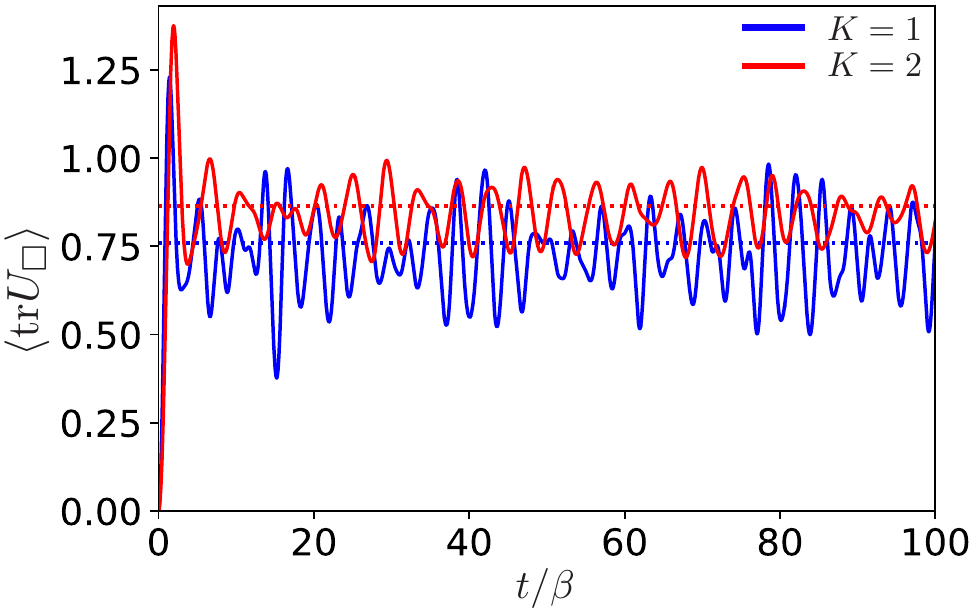}
    \includegraphics[width=.41\textwidth]{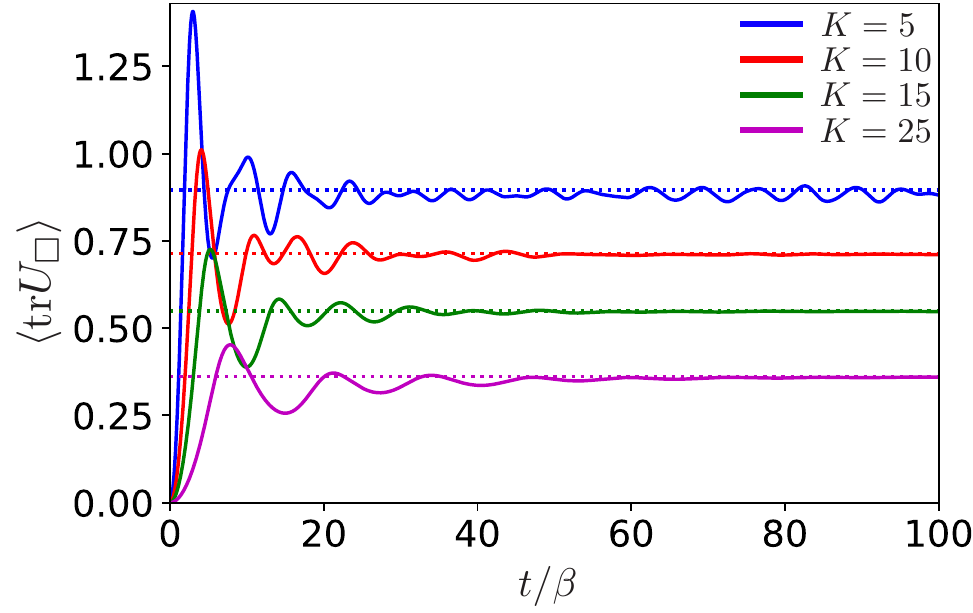}
     \caption{Time evolution of the Wilson loop defined on the plaquette colored in Fig.~\ref{fig1} in units of $\beta$ for 
     $K=1$ (top, blue), $K=2$ (top, red), 
     $K=5$ (bottom, blue), $K=10$ (bottom, red), $K=15$ (bottom, green), and $K=25$ (bottom, magenta). 
         The dashed lines show the canonical ensemble average. The values of $\beta$ are shown in Table~\ref{table1}.}
    \label{fig3}
    \end{figure}
    \begin{table}[tb]
      \begin{tabular}{c|cccccc}
        $K$ &1 & 2 & 5 &  10  &  15  &  25   \\
        \hline  
             $\overline{\langle\tr U_{\square}\rangle}$& $0.726$ & $0.842$ & $0.883$ &  $0.711$  &  $0.548$  &  $0.358$  \\
             $\langle\tr U_{\square}\rangle_{\rm can}$& $0.759$ & $0.865$ & $0.896$ &  $0.714$  &  $0.549$  &  $0.361$ \\
             $\beta$ &$0.937$ & $0.491$& $0.196$ & $0.0800$ & $0.0422$ & $0.0171$\\ 
             $\sigma$ &$0.15$& $0.075$& $0.014$& $0.0014$& $0.0014$& $0.0047$\\
             $\tau_{\rm eq}/\beta$ &$9.5$ & $6.4$& $6.4$ & $6.4$ & $6.6$ & $7.4$ \\
             $\tau_{\rm LE}/\beta$ &- & -& - & $0.23$ & $0.28$ & $0.40$ 
      \end{tabular}
      \caption{Time average after thermalization $\overline{\langle\tr U_{\square}\rangle}$, the canonical ensemble average $\langle\tr U_{\square}\rangle_{\rm can}$ of the Wilson loop, and the corresponding inverse temperature $\beta$. 
      $\sigma$,  $\tau_{\rm eq}$, and $\tau_{\rm LE}$ are
      the normalized standard deviation $\sqrt{\overline{(\langle\tr U_{\square}\rangle/\overline{\langle\tr U_{\square}\rangle}-1)^2}}$,
       the relaxation time to the thermal state, and the time scale of the exponential decay of the Loschmidt echo, respectively.
      }
	\label{table1}
    \end{table}
\subsection{Thermalization time}\label{sec:Numerical results}
We show the time evolution of the Wilson loop after the interaction quench in Fig.~\ref{fig3}.
We clearly see the Wilson loop rapidly reaches some equilibration value and fluctuates around it.
The fluctuation around the long time average decreases as $K$ increases, which is less than $1\%$ for $K>5$, while it is about $15\%$ for $K=1$ (see Table~\ref{table1}).
To reveal whether the Wilson loop reaches the thermal state or not, we computed the canonical ensemble average of the same Wilson loop operator, $({\rm Tr}\, e^{-\beta H}\tr U_{\square})/Z$, where ${\rm Tr}$ represents the trace over the physical Hilbert space, $\beta=1/T$ is the inverse temperature, and $Z={\rm Tr} e^{-\beta H}$ is the partition function.
We choose $\beta$ so that  the canonical average of the Hamiltonian $({\rm Tr} e^{-\beta H}H )/Z$ equals the expectation value in real-time evolution $\langle\Psi(t)|H|\Psi(t)\rangle=\langle\Psi(0)|H|\Psi(0)\rangle=E$, where $E$ is the total energy that is equal to zero in our state.
We found that the long-time average is in accord with the canonical ensemble average. 
This implies the Wilson loop gets equilibrated to the thermal state.
Since the fluctuations for $K=1,2$ are not small, we focus on the time evolutions for $K=5,10,15,25$ in the following.

\begin{figure}[t]
\centering
 \includegraphics[width=.41\textwidth]{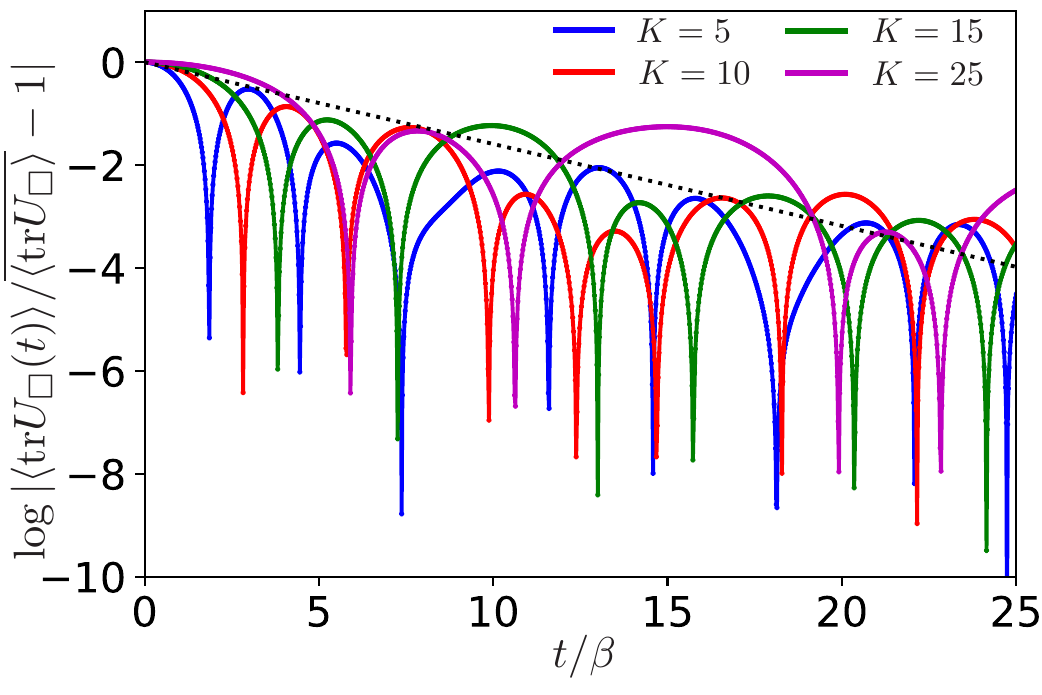}
 \caption{Normalized deviation from the long-time average in units of $\beta$, $\log|{\langle\tr U_{\square}(t)\rangle}/\overline{\langle\tr U_{\square}\rangle}-1|$,  for $K=5$ (blue), $K=10$ (red), $K=15$ (green), and $K=25$ (magenta). 
The values of $\beta$ are shown in Table~\ref{table1}.
 The black dotted-line shows $-t/(2\pi\beta)$. }
\label{fig4}
\end{figure}

Let us evaluate  the time scale of relaxation to the thermal equilibrium.
We show the log plot of the deviation from the long-time average in Fig.~\ref{fig4}.
The log plot shows linear decreasing behavior in time, 
which implies the exponential damping of the deviation,
${\langle\tr U_{\square}(t)\rangle}-\overline{\langle\tr U_{\square}\rangle}\sim e^{-t/\tau_{\rm eq}}$, with $\overline{\langle\tr U_{\square}\rangle}$ being the time average after thermalization.
Although the strength of the expectation values strongly depends on $K$, the time scale of thermalization is insensitive.
There is an ambiguity to determine the thermalization time due to oscillations of expectation values.
We, here, employ a linear fit using the peaks of the normalized deviation $\log|{\langle\tr U_{\square}(t)\rangle}/\overline{\langle\tr U_{\square}\rangle}-1|$
from $t=0$ to $t=25\beta$ in Fig.~\ref{fig4}. The time scales of thermalization are summarized in Table~\ref{table1},
which are typically $\tau_\text{eq}\sim6.5\times\beta\sim 2\pi \beta=2\pi/T$.
The time scale $2\pi/T$ is known as the Boltzmann time~\cite{Goldstein_2015,Reimann2016}.

For typical temperature of the quark-gluon-plasma produced in RHIC, $T=200$ MeV, $\tau_\text{eq}\sim 2\pi/T$ is $6$ fm/c.
This is an order of magnitude larger than the time scale expected from the hydrodynamic model ($\sim 0.5$ fm/c), and the thermalization time scale $1/(\pi T)\sim 0.3$fm/c, observed in calculations based on the gauge/gravity duality~\cite{Horowitz:1999jd,Chesler:2009cy,Heller:2011ju,Heller:2012km}.

\begin{figure}[t]
\centering
 \includegraphics[width=.42\textwidth]{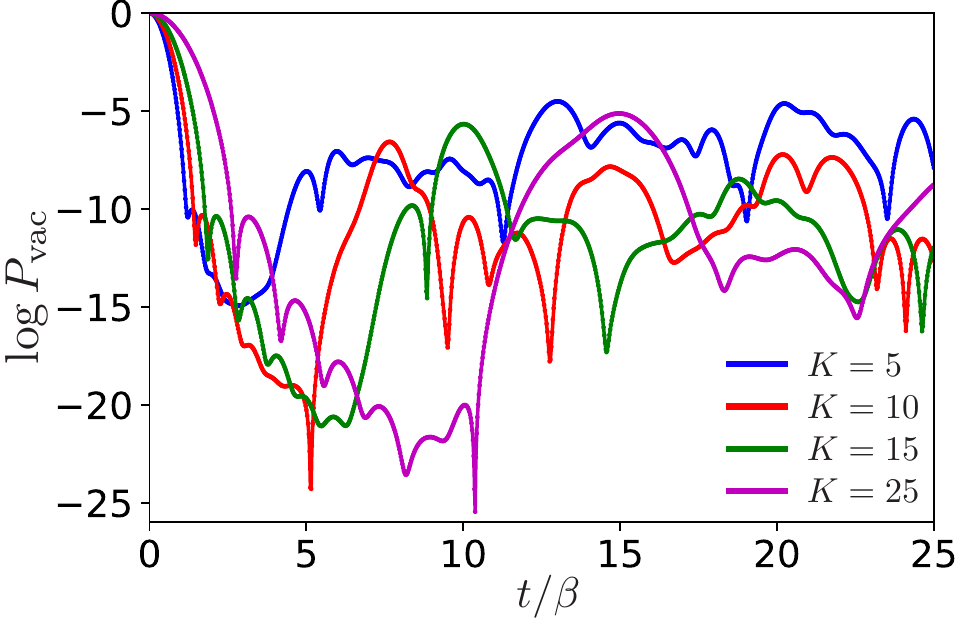}
 \caption{Time evolution of the logarithm of the vacuum persistent probability in units of $\beta$ for $K=5$ (blue), $K=10$ (red), $K=15$ (green), and $K=25$ (magenta). 
 The values of $\beta$ are shown in Table~\ref{table1}.}
\label{fig5}
\end{figure}

Next, to deepen understanding of the thermalization from the dynamics of the wave function, we compute the vacuum persistent probability, which is also known as the Loschmidt echo or fidelity in the context of quantum chaos~\cite{GORIN200633} and dynamical quantum phase transition~\cite{Heyl_2018}.
The Loschmidt echo is defined as
$P_{\rm vac}(t) =|\langle \Psi(0)|\Psi(t)\rangle|^2$,
and quantifies the deviation of the state at time $t$ [$|\Psi(t)\rangle$] from the initial state $t=0$ [$|\Psi(0)\rangle$]. 
We show the time evolution of the logarithm of the Loschmidt echo after the interaction quench in Fig.~\ref{fig5}.
The wave function rapidly spreads out to the entire physical Hilbert space, 
and then after the time scale where the Wilson loop gets equilibrated, the spreading also stops and fluctuates around equilibrium values.
We can see three characteristic time regions: early, intermediate, and late time.
At the early time,  the logarithm of the Loschmidt echo shows a quadratically decreasing, $\log P_{\rm vac}(t)\simeq  -\eta^2t^2$.
The value of $\eta$ depends on the strength of the interaction. We can nicely fit the data as $\eta \sim 2.5\times K$, which is independent of the temperature.
At the intermediate time, $\log P_{\rm vac}(t)$ is linearly damping with oscillations, $-t/\tau_{\rm LE}$.
Again, this is oscillating, so that we employ a linear fit using points at the peak positions. It is, however, not easy to evaluate $\tau_{\rm LE}$ for $K\leq 5$ because no clear peaks are found. We here only evaluate $\tau_{\rm LE}$ for $K>5$, which 
leads to the typical time scale, $\tau_{\rm LE}\sim 0.3\times\beta \sim 1/(\pi T) $.
This is comparable to the thermalization time scale observed in calculations based on the gauge/gravity duality~\cite{Horowitz:1999jd,Chesler:2009cy,Heller:2011ju,Heller:2012km}.
At the late time, $\log P_{\rm vac}(t)$ fluctuates around $-10$.

 \begin{figure}[t]
 \centering
 \includegraphics[width=.4\textwidth]{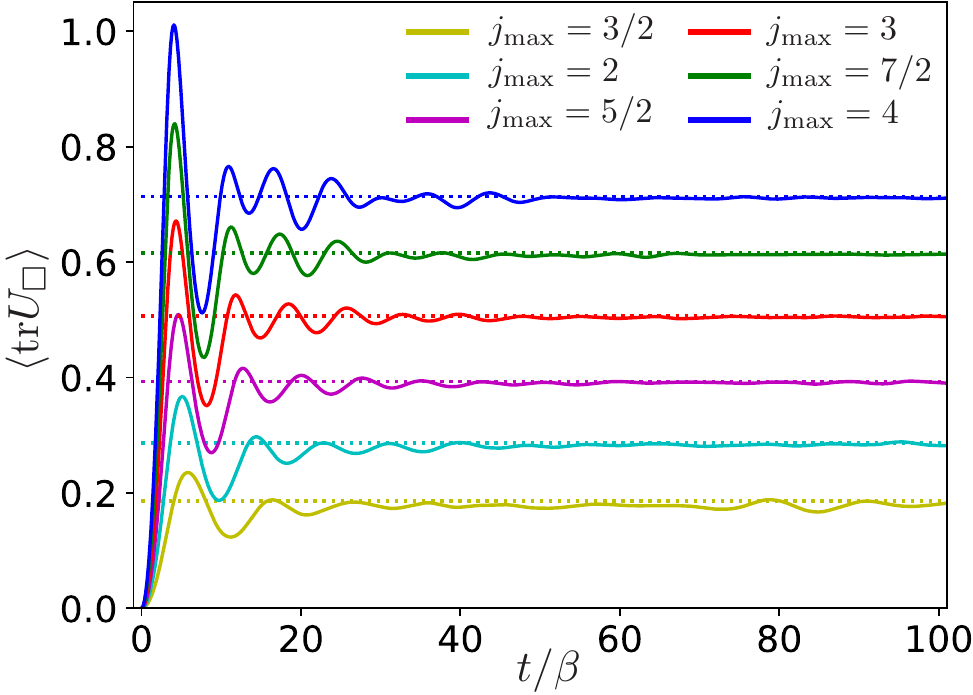}
 \caption{Time evolution of the Wilson loop defined on the bottom plaquette in Fig.~\ref{fig1}  in units of $\beta$ for $K=10$, and $j_{\rm max}=3/2$ (yellow), $j_{\rm max}=2$ (cyan),  $j_{\rm max}=5/2$ (magenta), $j_{\rm max}=3$ (red), $j_{\rm max}=7/2$ (green), and $j_{\rm max}=4$ (blue). 
 The values of $\beta$ for $K=10$, and $j_{\rm max}=3/2$ (yellow), $j_{\rm max}=2$ (cyan),  $j_{\rm max}=5/2$ (magenta), $j_{\rm max}=3$ (red), $j_{\rm max}=7/2$ (green), and $j_{\rm max}=4$ (blue) are $0.03034$, $0.04093$, $0.05127$, $0.06173$, $0.07134$, and $0.07996$, respectively.
     The dashed lines show the canonical ensemble averages.}
 \label{fig6}
 \end{figure}
\begin{figure}[t]
  \centering
   \includegraphics[width=.41\textwidth]{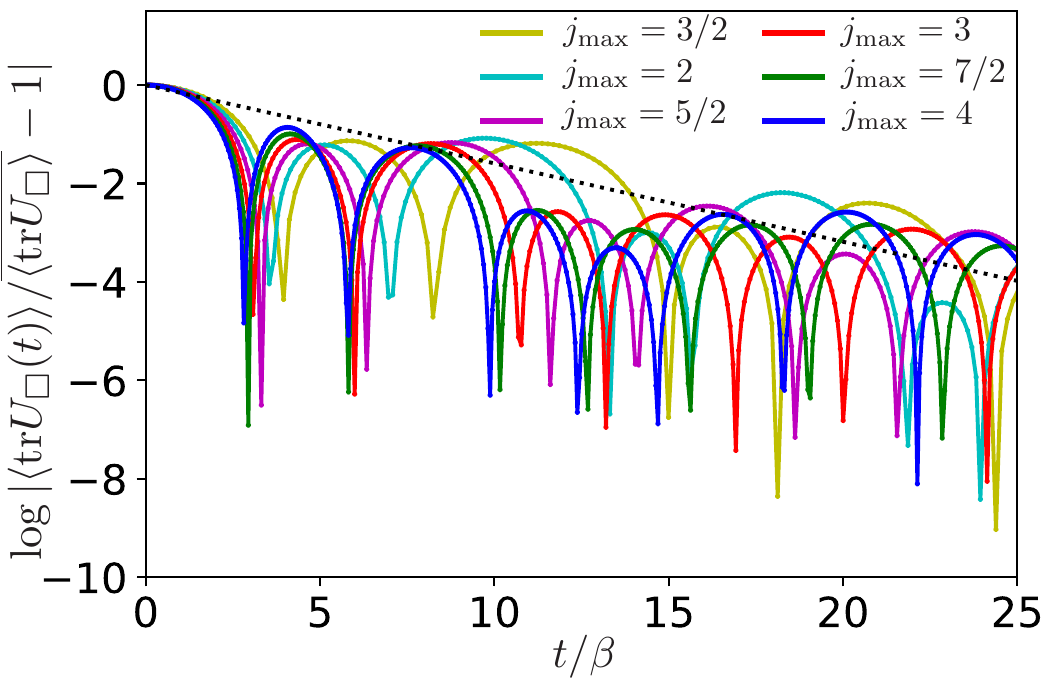}
   \caption{Normalized deviation from the long-time average in units of $\beta$, $\log|{\langle\tr U_{\square}(t)\rangle}/\overline{\langle\tr U_{\square}\rangle}-1|$,  for $K=10$ and $j_{\rm max}=3/2$ (yellow), $j_{\rm max}=2$ (cyan),  $j_{\rm max}=5/2$ (magenta), $j_{\rm max}=3$ (red), $j_{\rm max}=7/2$ (green), and $j_{\rm max}=4$ (blue).
   The black dotted line shows $-t/(2\pi\beta)$. }
  \label{fig7}
  \end{figure}
  \begin{figure}[!t]
 \centering
 \includegraphics[width=.43\textwidth]{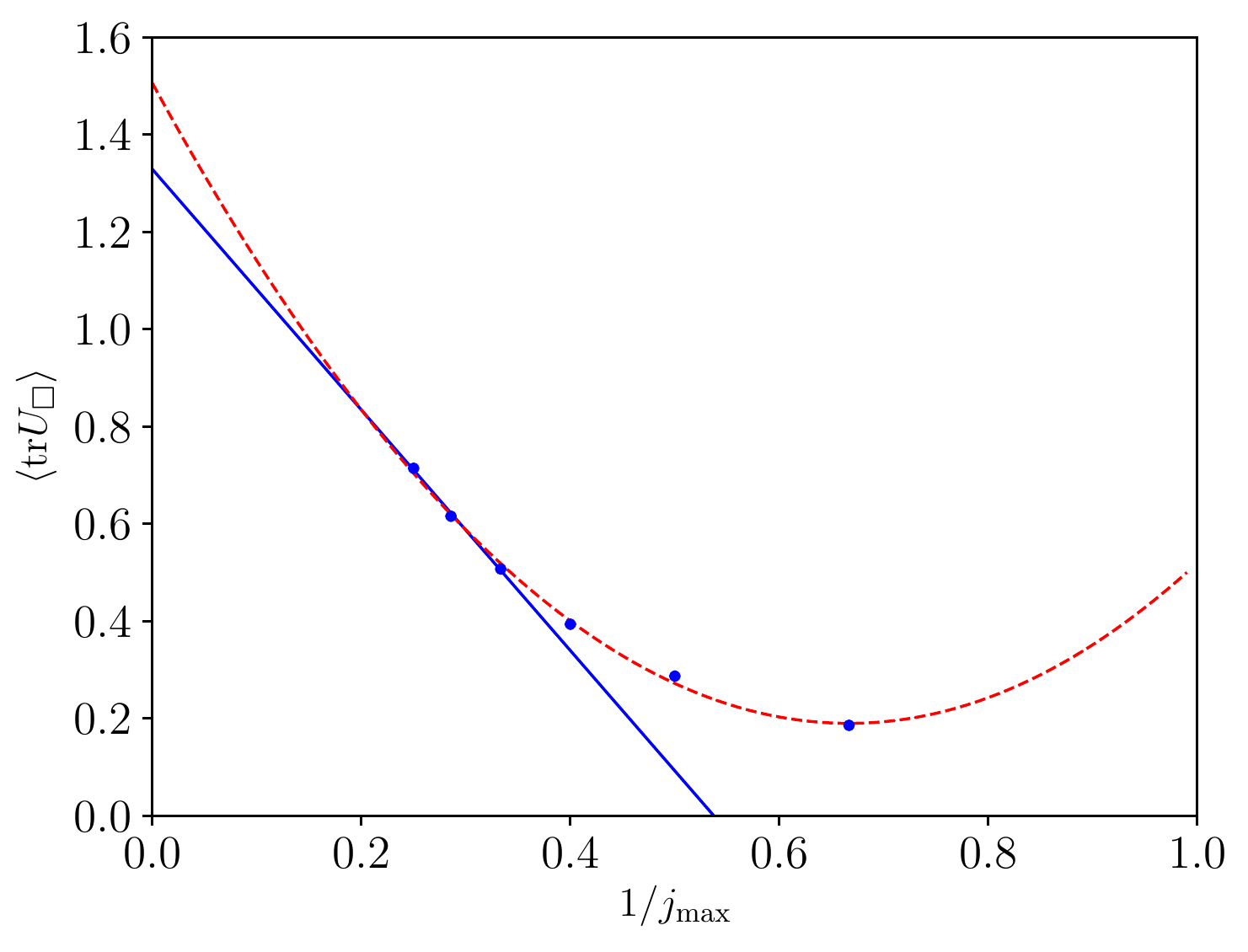}
 \caption{Extrapolation of the canonical average of the Wilson loop to $j_{\rm max}\rightarrow\infty$ for $K=10$. 
Blue dots show the canonical average of the Wilson loop at each $j_{\rm max}$.
Blue solid and red dashed curves show the results of the linear and quadratic polynomial fittings, respectively.}
 \label{fig8}
 \end{figure}

\subsection{$j_{\rm max}$ dependence}
Finally to see the dependence of the truncation of spins (i.e., the $j_{\rm max}$ dependence), we show the time evolution of the Wilson loop by changing $j_{\rm max}$ in Fig.~\ref{fig6} with an intermediate coupling $K=10$.
We also show the $j_{\rm max}$ dependence of the normalized deviation of the same Wilson loop from its long-time average in Fig.~\ref{fig7}. 
We see that the relaxation time scale is insensitive to the choice of $j_{\rm max}$, in particular for $j_{\rm max}>2$, where the dimension of the Hilbert space is larger than $10^{6}$.
In this case, the quantitative discussion of the relaxation time is possible within the numerically reachable $j_{\rm max}$.

To see the $j_{\rm max}$ dependence of the absolute value of the Wilson loop.
We show the canonical average of the Wilson loop (corresponding to stationary values in Fig.~\ref{fig6}) by changing $j_{\rm max}$ in Fig.~\ref{fig8} with an intermediate coupling $K=10$.
Using the linear and quadratic polynomial fitting, we estimate the canonical average with $j_{\rm max}\rightarrow\infty$ as $1.33\pm0.08$ (linear), and $1.50\pm0.13$ (quadratic).
The largest three $j_{\rm max}$ are used for the linear fitting, while all $j_{\rm max}$ are used in the quadratic fitting.
Here the errors are estimated from the $95$\% confidence interval. We show the fitting curves in Fig.~\ref{fig8}. 
Although the two estimations give the consistent results within the error bars, each data strongly depends on $j_{\rm max}$ as seen in Fig.~\ref{fig8}, and results in large extrapolation errors.
This result implies that we may need larger $j_{\rm max}$ for the complete quantitative research.

\section{Summary and outlook}\label{sec:Summary}
We have studied the real-time evolution of the SU($2$) Yang-Mills theory in a $(3+1)$ dimensional small lattice system after interaction quench. We have numerically solved the Schr{\"o}dinger equation in the reduced Hilbert space obtained by explicitly solving the Gauss law constraints.
We have observed the thermalization to the canonical state; the relaxation time $\tau_{\rm eq}$ is insensitive to the strength of the coupling constant, and scaled by the Boltzmann time $2\pi/T$.
The observed thermalization is very rapid compared with conventional matters, but it is still an order of magnitude larger than the one expected from the hydrodynamic model.

We hope that our numerical simulations in small systems share essential features of nonequilibrium dynamics with real QCD,
although we need to confirm it by conducting a more comprehensive study in future works, e.g., checking the $j_{\rm max}$, system size,  and initial-state dependences of the relaxation time, changing the lattice geometry, generalizing to the SU($3$) group, and so on.
In particular, the strong $j_{\rm max}$ dependence is observed in the absolute value of the Wilson loop, although the relaxation time is less insensitive to $j_{\rm max}$. We may elaborate on these in future research.

Furthermore, using our formulation, we can attack important problems of nonequilibrium QCD.
For example, we can compute the Kubo formula and estimate transport coefficients in a small system.
We can also compute the so-called out of time-order correlators, and confirm whether the lattice Yang-Mills theory saturates the maximum bound on the quantum Lyapunov exponent conjectured on the basis of the gauge/gravity duality~\cite{Maldacena:2015waa}.

\begin{acknowledgments}
T.~H. thanks to T.~Doi and Y.~Kikuchi for useful comments.
This work was supported by JSPS KAKENHI Grant Numbers~17H06462 and 18H01211.
The numerical calculations were carried out on XC40 at YITP in Kyoto University, and on cluster computers at iTHEMS in RIKEN.
We used PETSc~\cite{petsc-web-page}, SLEPc~\cite{Hernandez:2005:SSF}, and their Python bindings (petsc4py and slepc4py~\cite{DALCIN20111124}) for high-performance computing of matrix multiplications and matrix exponentials.
\end{acknowledgments}
\appendix
\onecolumngrid
\section{Eigenvalues of the Hamiltonian}\label{sec:Eigenvalues of the Hamiltonian}
We show the eigenvalues of the Hamiltonian near the ground state in Table~\ref{table3}.
 
\begin{table}[htb]
  \begin{tabular}{c|cccccc}
    $K$ &1  & & 2 & & 5 &    \\
     &$-E_n$  & error$\times 10^{-10}$ & $-E_n$  & error$\times10^{-10}$ & $-E_n$  & error$\times10^{-10}$         \\
    \hline  
     $n=0$      & $3.469117$ & $2.79713$ & $10.605675$ & $3.88412$  &  $37.056150$  &  $54.1126$  \\
     $n=1$     & $1.629849$ & $12.5664$ &  $7.448886$ & $99.9802$  & $31.451932$  &  $3.23442$  \\
     $n=2$    & $1.130174$ & $7.8021$ & $7.053874$ &  $51.3475$ & $31.133443$  &  $0.189302$ \\
      $n=3$   & $1.130174$ & $11.1765$ & $7.053874$ &  $3.1379$  &  $31.133443$  &  $35.0567$  \\
      $n=4$    & $1.093863$ & $53.1296$ & $6.880610$ &  $0.431258$  &  $31.012195$  &  $1.14193$ \\
      $n=5$    & $1.093863$ & $14.877$ & $6.872686$ &  $0.326881$  &  $31.012195$  & $9.3152$  \\
       $n=6$   & $0.831139$ & $12.3661$ & $6.418108$ &  $0.365907$  &  $30.694136$  & $5.33815$  \\
       $n=7$   & $0.831139$ & $5.87138$ & $6.045263$ &  $64.4065$  &  $30.694136$  &  $9.80712$ \\
       $n=8$   & $0.501544$ & $15.4608$ & $5.975238$ &  $70.7649$  &  $30.296693$  &  $4.33372$ \\
       $n=9$   & $0.357165$ & $46.6869$ & $5.360013$ &  $74.6421$  &  $29.662805$  &  $1.00764$  \\
    \hline  
            \hline  
      $K$ &  10 & &   15  & &  25 &   \\
     &  $-E_n$  & error$\times10^{-10}$ &   $-E_n$  & error$\times10^{-10}$ &  $-E_n$  & error$\times10^{-10}$   \\
    \hline  
     $n=0$        & $86.314621$ & $46.0226$ & $137.984803$ & $82.529$ &$244.122901$ & $0.232491$\\
     $n=1$      & $77.953834$ & $20.0109$ & $127.199330$ & $13.5481$ & $229.047852$ & $7.11994$\\
     $n=2$     &$77.712523$ & $85.9999$ & $127.152864$ & $10.0076$ & $228.669146$& $0.356782$\\
      $n=3$    &$77.601997$ & $0.714614$ & $127.000351$ & $18.6038$ & $228.577679$ & $5.48053$\\
      $n=4$     &$76.931297$ &$4.60878$ &$125.941700$ & $51.8256$ & $227.116557$ & $1.02326$\\
      $n=5$    & $76.594731$ & $1.61318$ & $125.776843$ & $24.9709$ & $226.648812$ & $7.05105$\\
       $n=6$     & $75.653934$ & $2.75647$ & $124.554066$ & $10.3648$ &$225.415640$ & $11.0096$ \\
       $n=7$   & $75.617178$ & $3.59361$ & $124.500679$ & $7.46896$ & $224.844530$ & $30.2425$ \\
       $n=8$    & $73.421128$ & $12.9538$ & $121.837617$ & $23.7776$& $221.925541$ & $4.6232$\\
       $n=9$   & $72.337828$ & $42.278$ & $120.507650$ &  $93.8186$ & $220.133792$ & $23.9231$\\
  \end{tabular}
  \caption{Smallest ten eigenvalues of the Hamiltonian and the relative error $||(H-E_n)|\psi_n\rangle||/||E_n|\psi_n\rangle||$ with $|\psi_n\rangle$ being the eigenvector for $j_{\rm max}=4$.}
\label{table3}
\end{table}

\bibliographystyle{apsrev4-1}
\bibliography{./gauge}

\end{document}